\documentclass[aps,prl,twocolumn,showpacs,amsmath,amssymb,amsfonts,nofootinbib,long]{revtex4}
\usepackage{graphicx}

\def\bef{\begin{figure}}
\def\eef{\end{figure}}

\newcommand{\bd}{\begin{displaymath}}
\newcommand{\ed}{\end{displaymath}}

\def\lsim{\raise0.3ex\hbox{$\;<$\kern-0.75em\raise-1.1ex
          e\hbox{$\sim\;$}}}
\def\gsim{\raise0.3ex\hbox{$\;>$\kern-0.75em\raise-1.1ex
\hbox{$\sim\;$}}}

\def\simlt{\mathrel{\lower2.5pt\vbox{\lineskip=0pt\baselineskip=0pt
           \hbox{$<$}\hbox{$\sim$}}}}
\def\simgt{\mathrel{\lower2.5pt\vbox{\lineskip=0pt\baselineskip=0pt
           \hbox{$>$}\hbox{$\sim$}}}}
\def\unity{{\hbox{1\kern-.8mm l}}}

\newcommand{\gr}{\mathbf}

\newcommand{\br}{\langle}
\newcommand{\kt}{\rangle}
\newcommand\G{\mbox{G}}
\newcommand\Mpc{\mbox{Mpc}}
\newcommand\eV{\mbox{eV}}
\newcommand\keV{\mbox{keV}}
\newcommand\MeV{\mbox{MeV}}
\newcommand\GeV{\mbox{GeV}}

\newcommand\CMB{\mbox{\tiny{CMB}}}
\newcommand\BBN{\mbox{\tiny{BBN}}}

\newcommand\B{\gr{B}}
\newcommand\E{\gr{E}}

\newcommand\f{f_{12}}

\def\de{\partial}

\begin{document}

\title{Production of Axions by Cosmic Magnetic Helicity}
\author{L. Campanelli$^{1,2}$}
\email{campanelli@fe.infn.it}
\author{M. Giannotti$^{1,2}$}
\email{giannotti@fe.infn.it}

\affiliation{$^{1}${\it Dipartimento di Fisica, Universit\`a di Ferrara, I-44100 Ferrara, Italy
\\           $^{2}$INFN - Sezione di Ferrara, I-44100 Ferrara, Italy}}

\date{April, 2006}


\begin{abstract}
We investigate the effects of an external magnetic helicity production
on the evolution of the cosmic axion field. It is shown that a
helicity larger than $(few \times 10^{-15} \G)^2 \Mpc$, if produced at
temperatures above a few $\GeV$, is in contradiction with the
existence of the axion, since it would produce too much of an axion
relic abundance.
\end{abstract}


\pacs{14.80.Mz, 98.62.En} \maketitle

Recently, the topic of cosmic magnetic helicity generation and
phenomenology has been widely discussed (see, e.g.,
\cite{Helicity,Fie92,Magnetic,Grasso01,Sok05,noi,Tina1,Tina2}). An
in depth study of this very peculiar quantity could help in
understanding the nature of the cosmic magnetic field itself and
its generation. Magnetic helicity is defined as
\begin{equation}
\label{helicity}
    H_B(t) = \frac1V
    \! \int_V \, \! d^3 x \, {\textbf A} \cdot \nabla \times {\textbf
    A},
\end{equation}
where $A^{\mu} = (A^0,{\textbf A})$ is the electromagnetic field.
In a flat universe described by a Robertson-Walker metric, $ds^2 =
dt^2 - R^2 d{\textbf x}^2$, where $R(t)$ is the expansion
parameter normalized so that at the present time $t_0$, $R(t_0)
=1$, the electric and magnetic fields are defined as ${\textbf E}
= -R^{-1} \dot{{\textbf A}}$ and $ {\textbf B} = R^{-2} \nabla
\times {\textbf A}$, where a dot indicates the derivative with
respect to the cosmic time $t$. Hence, the helicity can be written
as $H_B(t) = (1/V) R^2 \! \int_V \, \! d^3 x \, {\textbf A} \cdot
{\textbf B}$. It should be noted that in the literature the
definition of $H_B$ is usually given without the factor $R^2$. In
any case, the two definitions coincide at the present time.
However we note that, since the early universe is a very good
conductor, magnetic fields are frozen into the plasma and then $B$
scales in time as $B \propto R^{-2}$ \cite{Magnetic}. Therefore,
magnetic helicity, as defined in Eq.~(\ref{helicity}), remains
constant (after its generation).

The magnetic helicity is related to the topological properties of
the magnetic field, it is a $CP-$odd pseudo-scalar quantity and,
if different from zero, would reveal a macroscopic $P$ and $CP$
violation in the universe.
\\
Let us introduce the magnetic energy and magnetic helicity spectra
${\mathcal E}_B(k,t) = (2\pi/V) k^2 \, {\textbf B}({\textbf k})
\cdot {\textbf B}({-\textbf k})$
and
${\mathcal H}_B(k,t) = (4\pi/V) R^2 k^2 {\textbf A}({\textbf k})
\cdot {\textbf B}({-\textbf k})$,
respectively, where ${\textbf B}({\textbf k})$ and ${\textbf
A}({\textbf k})$ are the magnetic field and the vector potential
in Fourier space and $k = |{\textbf k}|$. In terms of the spectra,
the magnetic energy, $E_B(t) = (1/2V) \int_V \! d^{\,3} x \,
{\textbf B}^2({\textbf x},t)$, and the magnetic helicity are
$E_B = \int \! dk \, {\mathcal E}_B$, and
$H_B = \int \! dk \, {\mathcal H}_B$.
From the above definitions it follows that any magnetic field
configuration satisfies the inequality
$|{\mathcal H}_B (k,t)| \leq 2 R^4 k^{-1} {\mathcal E}_B (k,t)$
\cite{noi}.
A straightforward integration in $k$ leads to the so-called
``realizability'' condition, $|H_B| \leq H_B^{max} = 2R^3 \xi_B
E_B$, which fixes the maximal helicity associated with a given
magnetic field configuration. Here,
$\xi_B(t) = 2 \pi R \int \! dk \, k^{-1} {\mathcal E}_B(k,t)/E_B$
is the magnetic field correlation length. Therefore, bounds on the
magnetic field can be directly translated into bounds on magnetic
helicity. Before discussing this point more deeply, it is useful
to introduce the following parametrization for the helicity
produced at the temperature $T_*$:
\begin{equation}
\label{max}
H_B(T_*) = \pm 3.3 \times10^{-19} \, r \:
\frac{g_*(T_*)^{1/2}}{g_{*S}(T_*)} \; \frac{\GeV}{T_*} \; \G^2
\Mpc ,
\end{equation}
where $g_{*}$ and $g_{*S}$ count the total numbers of effectively
massless degrees of freedom referring to the energy and entropy
density of the universe, respectively. (From now on, $g_{*}$ and
$g_{*S}$ will be considered equal.) The choice $r=1$ gives the
maximal helicity associated with a magnetic field correlated on
the Hubble scale, $\xi_B = H^{-1}$ ($H$ is the Hubble parameter),
and whose energy density equals the total radiation energy density
at $T_*$. Thus, the parameter $r\leq 1$ measures the fraction of
helicity with respect to the maximum possible at a certain time.

It is known from the Big Bang Nucleosynthesis (BBN) argument that
a magnetic field $B$ at the temperature $T \simeq 0.1 \MeV$,
correlated on the Hubble scale at that time, cannot be more
intense than $10^{11} \G$ \cite{Grasso01}. This yields the
upper limit
$r\lesssim r_{\BBN} \simeq 82 \, g_*(T_*)^{1/2}\, (T_*/\GeV)$.
On the other hand, the analysis of the observed anisotropy of the
Cosmic Microwave Background (CMB) radiation requires
the bound $10^{-9} \G$ on the scales of $1
\Mpc$ \cite{Grasso01} for the
magnetic field (today), which leads to
$r\lesssim r_{\CMB} \simeq 0.2 \, g_*(T_*)^{1/2}\,(T_*/\GeV)$.
%
Observe that for magnetic helicity produced before $1\GeV$ neither of
these limits give any interesting constraints, since $r$ is
always expected to be less than $1$.

In this paper we consider the effects of magnetic helicity
generation on the cosmic axion field, showing how such generation could
considerably amplify the axion's expected relic abundance $\Omega_a$. In
particular, we will show that the request $\Omega_a \leq
\Omega_{{\rm matter}}$ sets, in most cases, a bound on $r$ of many
orders of magnitude smaller that the above $r_{\BBN}$ and
$r_{\CMB}$.

The axion field $a$ emerges in the so called Pecce-Quinn (PQ)
mechanism~\cite{PQ} for the solution of the strong CP problem (for
a review see, e.g., ~\cite{kim}). In most models, it is identified
with the phase $\Theta$ of a complex scalar field $a = f_a
\Theta$, where $f_a$ is a phenomenological scale usually referred
to as the PQ- or axion-constant, and presently constrained in the
very narrow region $10^9 \lesssim f_a \lesssim 10^{12} \GeV$ by
astrophysical and cosmological considerations \cite{Raff}.
\footnote{For the so-called hadronic axion models, whose
interaction with electrons is suppressed, a small region around
$f_a \sim10^6 \GeV$ is also allowed \cite{hadronic,hadronic1}. For
other possibilities of reconsidering the limits on the PQ-constant
see, for example, Ref. \cite{Dav86}.}

For temperatures above the PQ-scale the total Lagrangian is
required to be invariant under a constant shift of the phase
$\Theta$ (PQ-symmetry). However, for lower temperatures, the
symmetry is (spontaneously) broken and the angle $\Theta$ is fixed
on a precise value. We indicate this initial value with
$\Theta_i$. This phase evolves following the equation \cite{kim}
\begin{equation}
\label{1} \ddot{\Theta} + 3H \dot{\Theta} + \frac{\de V}{\de
\Theta} = 0,
\end{equation}
where $V(\Theta,T)=m_a^2(T)(1-\cos{\Theta})$ is the instanton
induced potential. The temperature-dependent axion mass is
\cite{GPY,Kolb} $m_a(T) \simeq 0.1 m \left(\Lambda/T\right)^{3.7}$
for $T\gg \Lambda$ and $m_a(T) = m$ for $T \ll \Lambda$
with
$m \simeq 6.2 \times 10^{-6} \eV / \f$ [$\f \! = \! f_{a}/(10^{12}
\GeV)$],
and $\Lambda \! \sim \! 200 \, \MeV$ is the QCD scale.

As the curvature (mass) term in Eq.~(\ref{1}) becomes dominant over
the friction (Hubble) term, $\Theta$ begins to oscillate with the
frequency $m_a(T)$. This happens at about the temperature $T=T_1$
defined by the equation $m_a(T_1)=3 H(T_1)$. Approximately $T_1
\simeq 0.9 \f^{-0.175} \Lambda_{200}^{0.65} \, \GeV$ with
$\Lambda_{200} = \Lambda/(200 \, \MeV)$.
\\
During this period of coherent oscillations the number of axions
in a comoving volume remains constant, $n_a R^3=$~const, where
$n_a(T)$ is the axion number density. Thus, the axion relic
abundance today is
\begin{equation}
\label{Omega}
\Omega_a \simeq 0.2 \, \Lambda_{200}^{-0.65} \, [\Theta_1^2+ (
\dot\Theta_1/3H_1)^2 \, ] \, \f^{1.175},
\end{equation}
where $\Theta_1$, $\dot\Theta_1$ and $H_1$ are respectively
$\Theta(T_1)$, $\dot\Theta(T_1)$ and $H(T_1)$. If the axion field
is not interacting in the region $T>T_1$, and $T_1 \ll f_a$, then
the axion kinetic energy at $T=T_1$ is reliably negligible,
$\dot\Theta_1\propto(T_1/f_a)^3\ll 1$, and consequently
$\Theta_1\simeq\Theta_i$. Thus, Eq.~(\ref{Omega}) reduces to
$\Omega_{a} = \Omega_{a 0} \simeq 0.2 \, \Lambda_{200}^{-0.65}
\Theta_i^2 \f^{1.175}$.
Assuming the natural value $\Theta_i \sim1$, this gives
$\Omega_{a0} \simeq 0.3$ (the expected value for the dark matter
abundance) for $\f\sim1$. Much larger values of $\f$ would cause
too much axion production and are therefore excluded.

However, if the field evolution took place in the presence of an
external magnetic field, the previous analysis needs to be
modified. It is easy to show the importance of the magnetic
helicity in this case. In fact, in the presence of a magnetic
field, the right hand side of Eq.~(\ref{1}) is modified as
$(\alpha/2\pi f_a^2) \E \cdot \B$. Assuming a homogeneous axion
field, and observing that $(1/V) \int_V d^3 x \, \E \cdot \B = -
\dot H_B/(2R^{3})$, Eq.~(\ref{1}) becomes
\begin{equation}
\label{2} \ddot{\Theta} + 3H \dot{\Theta} + \frac{\de V}{\de \Theta}= -
\frac{\alpha}{4\pi f_a^2} \, \frac{\dot H_B}{R^{3}} \, .
\end{equation}
The slightness of the effects of a non-helical magnetic field on
the cosmic evolution of the axion field discussed in literature
(see, e.g., \cite{paradox}) can easily be understood: even though
the axion itself produces helicity in the presence of an external
magnetic field \cite{Fie92,noi,Lee00} and therefore its evolution
is modified, the effect is negligible since the helicity produced
by the axion itself is small. However, if another mechanism were
responsible for the production of a significant amount of
helicity, the consequences on the axion relic abundance today
could be enormous.
%

Let us assume, first, that a large amount of helicity is produced
at the temperature $T_* \gg T_1$ and that the time
scale of its production is short compared to the time
scale $t_1$.
\footnote{This is not a very strong assumption. For example, in
the $\alpha$-dynamo mechanism this time scale is expected to be
$\Delta t_* \lesssim t_* \ll t_1$ \cite{Sok05}.}
Moreover, taking into account that after its generation
the helicity can be considered a (quasi)-conserved quantity (see,
e.g., \cite{Tay74}), we can safely approximate it as a step
function, $H_B(t) = H_B(t_*) \, \theta(t-t_*)$.

When the potential energy is small compared to the kinetic energy
and/or the friction, Eq.~(\ref{2}) can be solved giving
$\dot\Theta(T) = \alpha H_{B}(T)/(4\pi f_a^2 R^{3})$,
that is, the kinetic energy is a decreasing function of time. Let
us indicate with $T_{\rm eq}$ the temperature at which the kinetic
energy equals the potential: $(\dot\Theta^2/2)|_{T_{\rm eq}}=\br
V(\Theta) \kt |_{T_{\rm eq}}$, where for the sake of simplicity we
are considering the root mean square value of the potential $\br
V(\Theta) \kt = m_a^2(T)(1-1/\sqrt{2})$. It is clear that only if
$T_{\rm eq}\geq T_1$ will the oscillations start at the
temperature $T_1$, where the friction becomes small. In this case,
the standard analysis applies and we find
\begin{equation}
\Omega_a\simeq 0.1 \f^{1.175} \Lambda_{200}^{0.65}
\left[1+\left(r/r_a\right)^2\right], \quad \mbox{if}~~ r\leq r_a,
\end{equation}
where
$r\leq r_a \simeq 1.5\times 10^{-11}
g_{*}^{1/2}(T_*)\Lambda_{200}^{-0.65}\f^{2.175}(T_*/\GeV)$
corresponds to $T_{\rm eq}\geq T_1$. However, if $T_{\rm eq}<T_1$
the oscillations cannot start at $T=T_1$, the kinetic energy being
still too large, but will be delayed and start at $T=T_{\rm eq}$.
The resulting axion relic abundance is
\begin{equation}
\Omega_a\simeq 0.2 \f^{1.175} \Lambda_{200}^{0.65}
\left(r/r_a\right), \quad \mbox{if}~~ r > r_a.
\end{equation}


\begin{figure}[t]
\begin{center}
\includegraphics[clip,width=0.43\textwidth]{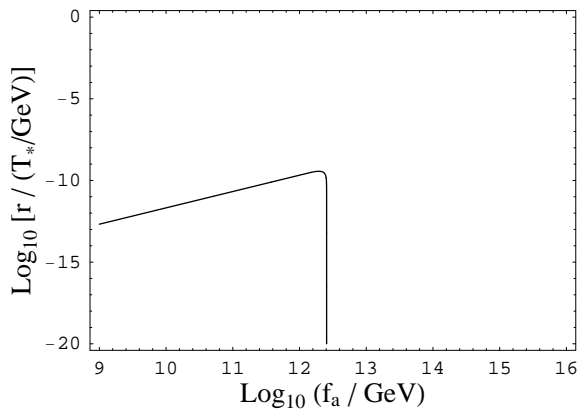}
\caption{Allowed region for the magnetic helicity and the
PQ-constant in the case of helicity production before the axion
coherent oscillations. The area under the curve represents the
parameter region $\Omega_a <\Omega_{{\rm matter}} \simeq 0.3$
\cite{WMAP}.} \label{fig:1}
\end{center}
\end{figure}


In Fig.~\ref{fig:1}, we show the cosmologically allowed region,
$\Omega_a <\Omega_{{\rm matter}} \simeq 0.3$, for the PQ-constant
$f_a$ and the $r$ parameter. Observe that, since we are now
considering helicity to be generated before the axion
oscillations, $T_* > T_1 \sim1 \GeV$, the BBN and CMB arguments do
not set any consistent limit on the magnetic helicity. On the
other hand, the limit on $r$ from the axion abundance is extremely
significant: the curve in Fig.~\ref{fig:1} is well approximated by
$r = r_a$ (for $\f \lesssim 1$) and this implies [see
Eq.~(\ref{max})] that a magnetic helicity generated at the
temperature $T_*$ cannot be larger than
\begin{equation}
\label{lim1} H_B \simeq 5 \times 10^{-30} \; \Lambda_{200}^{-0.65}
\f^{2.175} \: \G^2 \Mpc.
\end{equation}
In the most conservative case, $\f \simeq 1$, we exclude any
$H_B\gsim(few \times 10^{-15} \G)^2 \Mpc$.
\\
We note that a full numerical treatment of the axion evolution
equation would reveal the existence of a small region with $f_a$
above the cosmological bound. We omit the discussion of this case
to avoid useless complications. Whatever the case, the region
emerges as a fine tuned compromise between helicity and the
$\Theta$-angle, and is therefore very small.

\begin{figure}
\begin{center}
\includegraphics[clip,width=0.43\textwidth]{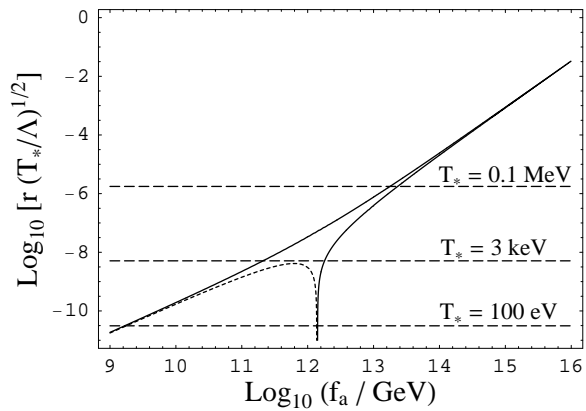}
\caption{Allowed region for the magnetic helicity and the
PQ-constant in the case of fast helicity production during the
axion coherent oscillations. The areas enclosed by the dashed and
continuous curves represent the parameter region $\Omega_a <
\Omega_{{\rm matter}} \simeq 0.3$ for the plus and minus sign in
Eq.~(\ref{Q}), respectively. The areas under the horizontal lines
represent the region allowed by the CMB analysis for three
different temperatures.} \label{fig:2}
\end{center}
\end{figure}

We will now briefly discuss the phenomenon of helicity production
during the axion oscillations and show that the effects on the
axion relic abundance are, in this case, less evident. The
peculiarity of this second case is the appearance of an additional
energy scale, the axion mass, which was neglected in the above
discussion. In the radiation era Eq.~(\ref{2}) can be re-stated as
$\ddot{\varphi} + (3H^2/4 + m_a^2) \varphi = F(t)$, where $\varphi
= \Theta R^{3/2}$, $F(t) = - (\alpha/4\pi f_a^2)(\dot
H_B/R^{3/2})$, and we assumed $|\Theta|<1$ (see below) so that
$V(\Theta)\simeq m_a^2\Theta^2/2$. For temperatures $T \ll
\Lambda$ the axion mass is constant, $m_a(T) = m$, and much larger
then the Hubble parameter, $H/m\sim 10^{-6}\Lambda_{200}^2\f\,
g_*(T)(T/\Lambda)^2$. Hence, factors of order $H/m_a$ can be
neglected in the previous equation, which therefore describes a
forced harmonic oscillator with frequency $m$ and forcing term
$F(t)$. Its solution is $\varphi = \varphi_0 + (1/m)\int_{t_i}^t
\! du \, F(u) \sin m (t-u)$, where $\varphi_0$ is the solution for
the free harmonic oscillator. With a suitable choice of initial
conditions we can write $\varphi_0 = A \sin(mt)$, $A$ being
related to the actual axion energy density in the case of null
magnetic helicity, $\rho_a(T_0) = \frac{1}{2} f_a^2 m^2 A^2$.
\\
Let us suppose that the helicity is produced during the finite
interval of time $(t_A,t_B)$. Since before $t_A$ and after $t_B$
the evolution of $\varphi$ is that of a free oscillator, and since
for $r\lesssim r_{\CMB}$ it results $|\Theta| < 1$ if $T_*
\lesssim 1 \MeV$, we can apply the standard procedure to find the
axion relic abundance today. This reads
%
%
$\Omega_a = \Omega_{a 0} + 2 \, \Omega_{a 0}^{1/2} \, \mbox{Re}z +
|z|^2$,
where
$z = \left(f_a/\sqrt{2 \rho_{c}}\right) \int_{t_A}^{t_B} \!\! dt
\, F(t) \, e^{imt},$
$\rho_{c}$ is the present critical density, and $\Omega_{a 0}$ is
the expected axion relic abundance (today) in the absence of
magnetic helicity [see the discussion below Eq.~(\ref{Omega})].
The time profile of the magnetic helicity in the expanding
universe is conveniently parameterized by the following
``rapidity'' function,
%
%
%
$f(t) = (R_*/R)^{3/2} \, t_{*} \, \dot{H}_B / H_B(t_*)$,
where $t_{*}$ is a reference time between $t_A$ and $t_B$.
Therefore, the forcing term can be written as $F(t) =
-(\alpha/4\pi f_a^2) [H_B(T_*)/R_{*}^{3/2}] f(t)$ and, using
Eq.~(\ref{max}), $z$ becomes
\begin{equation}
\label{Q} z \simeq \pm 3 \times 10^7 \Lambda_{200}^{1/2}
\,f_{12}^{-1} r\, (T_{*}/\Lambda )^{1/2}\, \hat{f}(m) \, ,
\end{equation}
%
%
where $\hat{f}(\omega) = \int_{-\infty}^{+\infty} dt f(t) \, e^{i
\omega t}$ is the Fourier transform of $f(t)$.
The axion relic abundance can be exactly calculated once the
rapidity function $f(t)$ is known.
In the case of slow ($\Delta t\gg m^{-1}$) helicity production it
results $\hat{f}(m) \sim 0$ and, consequently, $\Omega_a \simeq
\Omega_{a 0}$. On the other hand, if helicity production is fast
($\Delta t\ll m^{-1}$)
%
%
we have $|\hat{f}(m)| \sim \mbox{Re} [\hat{f}(m)] \sim 1$. To give
a simple estimate of $\Omega_{a}$ we observe that for $\f < 1$ it
results that $\Omega_{a 0} \ll \Omega_{a}$, and therefore
$\Omega_a\sim |z|^2$. This yields the upper bound
$r \lesssim 10^{-8} \Lambda_{200}^{1/2} \, \f
\left(\Lambda/T_*\right)^{1/2}$
which translates into a bound on the present magnetic helicity
\begin{equation}
\label{lim2} H_B \lesssim 10^{-27} \Lambda_{200}^{1/2} \,
g_*(T_*)^{-1/2} \f \! \left(\frac{\GeV}{T_*}\right)^{\!3/2} \!
\G^2 \Mpc.
\end{equation}
This limit is always stronger than the one from the BBN and, for
$T_* \gsim few \, \f^{2/3} \keV$, it also results that
$r<r_{\CMB}$ (see Fig.~\ref{fig:2}). In this figure, we have
included the region above the cosmological limit $f_a>10^{12}\GeV$
(see the continuous line), since the analysis is straightforward
in this case. Simply note that the minus sign in Eq.~(\ref{Q})
corresponds to a reduction of the axion relic abundance.


To conclude, we have pointed out that a strong connection between
magnetic helicity and axion cosmology exists. In particular, we
have studied the influence of primordial magnetic helicity
generation on the present axion relic abundance. The result is
that, allowing for possible helicity generation in the early
universe, the knowledge of the PQ-constant would not be sufficient
to determine the axion relic abundance today. Indeed, suppose
axions were detected with the PQ-constant strictly under $10^{12}
\GeV$. This would not exclude the possibility that they represent
the dark matter component of our universe, if an appropriate
amount of helicity has been generated (see Fig.~1). We focused our
analysis on two different cases corresponding to helicity
production before or during the axion coherent oscillations. Let
us consider the former, and more interesting, case first. We look
at two different possibilities within this case: {\it i}) first,
if axions were detected as cold dark matter particles this would
severely constrain the allowed amount of magnetic helicity in our
universe, according to Eq.~(\ref{lim1}). On the other hand, {\it
ii}) if a helicity greater than $(few\times 10^{-15} \G)^2\Mpc$
were revealed at the present time this would be in contradiction
with the existence of the invisible axion itself, unless it is
assumed that helicity production took place at a temperature above
the PQ-scale. It should be noted that the bounds found are
considerably stronger than the ones deduced from the BBN or CMB
analysis.

Finally, if helicity production took place during the
primordial axion oscillations, 
the constraints on the helicity depend on the time scale of its
production, and are indeed relevant only if this is much smaller
than the typical axion oscillation time $1/m_a$. Also in this
case, the bounds found, Fig.~\ref{fig:2} and Eq.~(\ref{lim2}), are
relevant compared to the ones from the BBN and CMB.

We consider these results of interest in view of next generation
experiments devoted to the axion search and future experiments on
CMB, which are expected to be sensitive enough for the detection
of a possible helical magnetic field \cite{Tina1,HelCMB}.


\begin{acknowledgments}
We would like to thank A. D. Dolgov for his careful reading of the
manuscript, and M. Giovannini and A. Mirizzi for helpful
discussions.
\end{acknowledgments}


\end{document}